\documentclass[]{article}
\usepackage[T1]{fontenc}
\usepackage[utf8]{inputenc}
\usepackage{authblk}
\usepackage{graphicx}
\title{Proper time and length in Schwarzschild geometry}
\author[1]{O. Brauer}
\author[2]{H. A. Camargo}
\author[3]{M. Socolovsky}
\affil[1,2]{Facultad de Ciencias, Universidad Nacional Aut\'onoma de M\'exico, Circuito Exterior, Ciudad Universitaria, 04510

 M\'exico D. F., M\'exico}
\affil[3]{Instituto de Ciencias Nucleares, Universidad Nacional Aut\'onoma de M\'exico, Circuito Exterior, Ciudad Universitaria,  04510

 M\'exico D. F., M\'exico}

\date{}

\begin{document}

\maketitle

\textbf{Abstract:} We study proper time ($\tau$) intervals for observers at rest in the universe ($U$) and anti-universe ($\bar{U}$) sectors of the Kruskal-Schwarzschild eternal spacetime of mass $M$, and proper lengths ($\rho$) in the black hole (BH) and white hole (WH) sectors. The fact that in asymptotically flat regions, coordinate time $t$ at infinity is proper time, leads to a past directed Kruskal time $T$ in $\bar{U}$. In the BH and WH sectors maximal proper lengths coincide with maximal proper time intervals, $\pi M$, in these regions, i.e. with the proper time of radial free falling (ejection) to (from) the singularity starting (ending) from (at) rest at the horizon.


PACS (2014): 04.70.Bw, 04.70.-s
\begin{center}
\line(1,0){350}
\end{center}

\section{Proper times in $U$ and $\bar{U}$}

All our analysis will be based on the Kruskal diagram corresponding to Schwarzschild spacetime i.e. in its maximal analytical extension \cite{hawking}. This is illustrated in Figure 1. We use geometric units $G_N=c=1$. 

\

In region $I=U$ (universe), $r>2M$, the square of the proper time is given by $$d\tau^2=(1-{{2M}\over{r}})dt^2-{{dr^2}\over{1-{{2M}\over{r}}}}-r^2d\Omega^2\eqno{(1)}$$ where $M$ is the mass and $d\Omega^2=d\theta^2+sin^2\theta d\varphi^2$. For $r\to\infty$ spacetime is flat and $t$ is the proper time. For an observer at rest at $r_0>2M$, $$d\tau^2=(1-{{2M}\over{r_0}})dt^2\eqno{(2)}$$ and therefore $$\Delta\tau_U=\sqrt{1-{{2M}\over{r_0}}}\Delta t, \ \ \Delta t=t_2-t_1.\eqno{(3)}$$ I.e. $\Delta\tau_U=\Delta\tau_U(M;r_0,\Delta t)$; so, $\Delta\tau_U\to 0$ as $r_0\to 2M$ (time ``does not pass" for light) and $\Delta\tau_U\to \Delta t$ as $r_0\to\infty$. The maximum value of $\Delta\tau_U$ is $+\infty$ since this is the maximum value of $\Delta t$. 

\

To determine the corresponding $\Delta\tau_{\bar{U}}$ (by symmetry it should equal $\Delta\tau_U$) we have to use the expression for $d\tau^2$ in terms of the Kruskal variables $T$ and $R$, valid in the four regions $U$, $BH$, $\bar{U}$, and $WH$ \cite{cheng}: $$d\tau^2=4\times{{2M}\over{r}} \ e^{-{{r}\over{2M}}} \ (dT^2-dR^2-r^2d\Omega^2),\eqno{(4)}$$ with $r$ implicitly given in terms of $T$ and $R$ by $${{1}\over{{2M}^2}}(R^2-T^2)=({{r}\over{2M}}-1)e^{{{r}\over{2M}}}.\eqno{(5)}$$ In region $I^\prime=\bar{U}$ (anti-universe) the relation between $T$ and $R$ with $t$ and $r$ is given by $$T=-2M\sqrt{{{r}\over{2M}}-1} \ e^{{{r}\over{4M}}} \ Sh({{t}\over{4M}})\in (-\infty,+\infty), \eqno{(6)}$$ and $$R=-2M\sqrt{{{r}\over{2M}}-1} \ e^{{{r}\over{4M}}} \ Ch({{t}\over{4M}})\in (-\infty,0). \eqno{(7)}$$ For constant $r$, $$dT={{\partial T}\over{\partial t}}dt=-{{1}\over{2}}\sqrt{{{r}\over{2M}}-1} \ e^{{{r}\over{4M}}} \ Ch({{t}\over{4M}})dt,$$ $$dR={{\partial R}\over{\partial t}}dt=-{{1}\over{2}}\sqrt{{{r}\over{2M}}-1} \ e^{{{r}\over{4M}}} \ Sh({{t}\over{4M}})dt$$ and therefore $$dT^2-dR^2=(({{\partial T}\over{\partial t}})^2-({{\partial R}\over{\partial t}})^2)dt^2={{1}\over{4}}({{r}\over{2M}}-1) \ e^{{{r}\over{2M}}} \ dt^2.\eqno{(8)}$$ Then, $d\tau^2_{\bar{U}}(r_0)=(1-{{2M}\over{r_0}})dt^2$ and so $$d\tau_{\bar{U}}(r_0)=\pm\sqrt{1-{{2M}\over{r_0}}}dt.\eqno{(9)}$$ The minus sign would lead to $d\tau_{\bar{U}}(r_0)<0$ for $dt>0$ or viceversa, $d\tau_{\bar{U}}(r_0)>0$ for $dt<0$. No of these results is admisible, since both $\tau$ (at finite distances) and $t$ (at infinity) are proper times, and {\it any proper time must always be future directed} \cite{synge}. Then, the plus sign has to be chosen in (9), and therefore $$\Delta\tau_{\bar{U}}(r_0)=\sqrt{1-{{2M}\over{r_0}}}\Delta t=\Delta\tau_U(r_0).\eqno{(10)}$$ But then $T_2<T_1$ i.e. {\it Kruskal time decreases}: $$\Delta T(r_0)=T_2-T_1=-2M\sqrt{{{r_0}\over{2M}}-1} \ e^{{{r_0}\over{4M}}} \ (Sh({{t_2}\over{4M}})-Sh({{t_1}\over{4M}})<0\eqno{(11)}$$ since $t_2>t_1$. The fact that $T$ is {\it past directed} in $\bar{U}$, indicates that $T$ is not a physical time in $\bar{U}$, but only a coordinate (though global) in Kruskal spacetime, with no intrinsic physical meaning, at least with respect to the eternal black hole. 

\begin{figure}[h!]
\hspace{-1cm}
\includegraphics[width=1.2\textwidth]{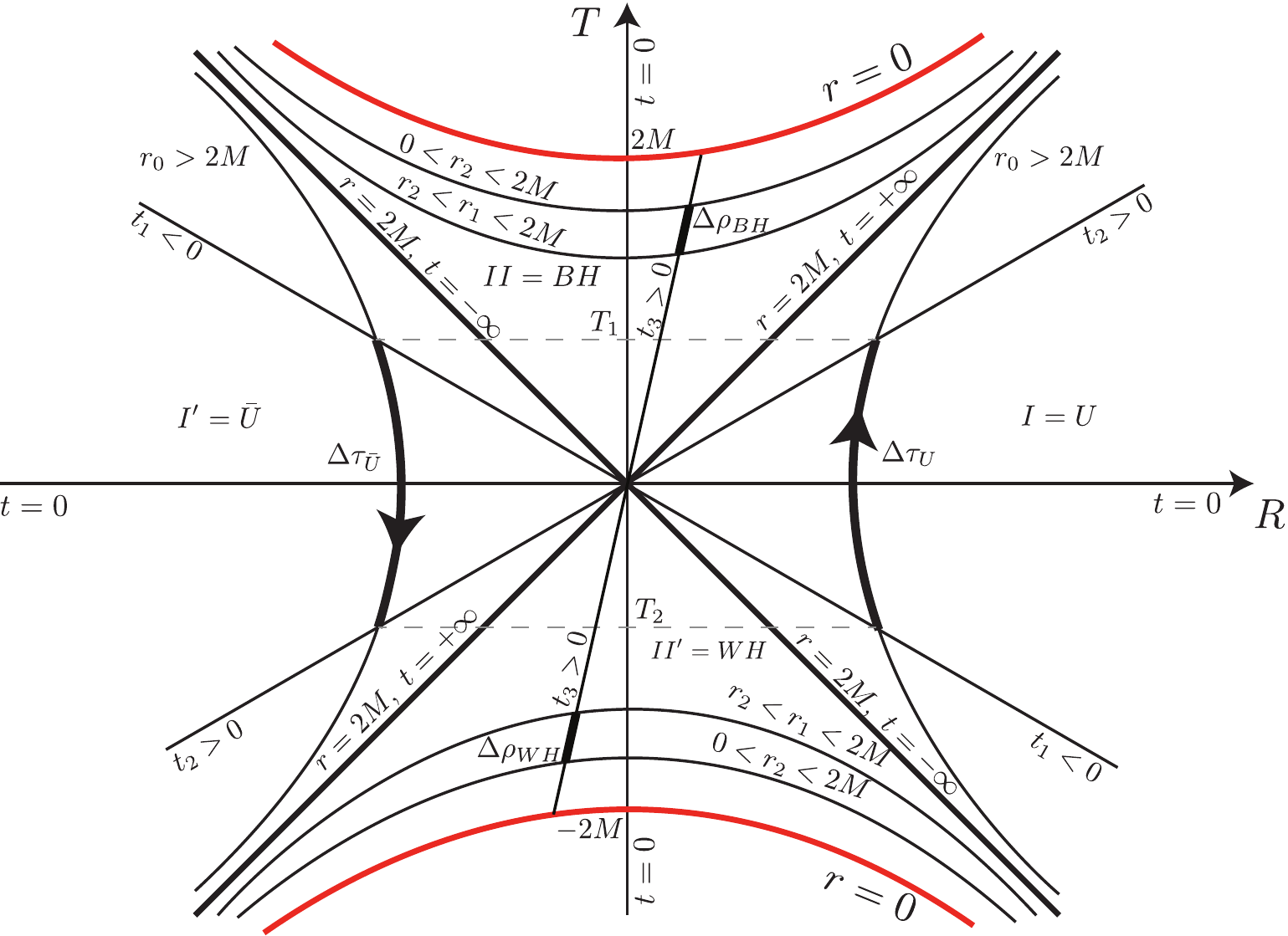}
\caption{Proper time and length in Kruskal diagram. }
\end{figure}

\section{Proper lengths in $BH$ and $WH$}

In region $II=BH$ (black hole), $0<r<2M$, the square of the interval is given by $$ds^2={{dr^2}\over{{{2M}\over{r}}-1}}-({{2M}\over{r}}-1)dt^2-r^2d\Omega^2\eqno{(12)}$$ which, at fixed $\theta$, $\varphi$, and time $t_3$ can be interpreted as the {\it elementary proper length} $d\rho$ along $dr$: $$d\rho_{BH}={{dr}\over{\sqrt{{{2M}\over{r}}-1}}}, \eqno{(13)}$$ independent of $t_3$. Integrating this expression between $r_2$ and $r_1$ gives $$\Delta\rho_{BH}(r_2,r_1)=\int_{r_2}^{r_1}dr\sqrt{{{r}\over{2M-r}}}=2M\int_{x_2}^{x_1}dx\sqrt{{{x}\over{1-x}}}\eqno{(14)}$$ with $x={{r}\over{2M}}$ and $x_i={{r_i}\over{2M}}$, $i=1,2$. Using $$\int dx\sqrt{{{x}\over{1-x}}}=-\sqrt{x(1-x)}+arctg({{\sqrt{x(1-x)}}\over{1-x}})+const. \eqno{(15)}$$ one obtains, in particular for the limits $r_2\to 0_+$ and $r_1\to (2M)_-$, $$\Delta\rho_{BH}(0,2M)=2\pi\times {{\pi}\over{2}}=\pi M.\eqno{(16)}$$ So, the {\it maximal proper length} in $BH$ coincides with the proper time of radial free falling to the future singularity at $r=0$ of a massive test particle, starting at rest from the future horizon \cite{raine}. 

\

By symmetry, the same result should hold in region $II^\prime=WH$ (white hole), but now the maximal proper length should coincide with the proper time of radial free ejection from the past singularity at $r=0$ of a massive test particle, ending at rest at the past horizon. In fact, in $II^\prime$ the relation between $T$ and $R$ with $t$ and $r$ is given by $$T=-2M\sqrt{1-{{r}\over{2M}}} \ e^{{{r}\over{4M}}} \ Ch({{t}\over{4M}})\in (-\infty,0), \eqno{(17)}$$ and $$R=-2M\sqrt{1-{{r}\over{2M}}} \ e^{{{r}\over{4M}}} \ Sh({{t}\over{4M}})\in (-\infty,+\infty). \eqno{(18)}$$ For constant $t$, $$dT={{\partial T}\over{\partial r}}dr={{r}\over{4M}} \ {{e^{{{r}\over{4M}}}}\over{\sqrt{1-{{r}\over{2M}}}}} \ Ch({{t}\over{4M}})dt,$$ $$dR={{\partial R}\over{\partial r}}dr={{r}\over{4M}} \ {{e^{{{r}\over{4M}}}}\over{\sqrt{1-{{r}\over{2M}}}}} \ Sh({{t}\over{4M}})dr,$$ and using again (4) with $d\theta=d\varphi=0$, one obtains $$d\rho_{WH}=d\rho_{BH}\eqno{(19)}$$ which, after integration, leads to the same results (14) and (16), but for $\Delta\rho_{WH}$. 

\section{Final comment}

It is believed that, probably, eternal black holes do not exist in nature, and that only black holes resulting from gravitational collapse (and also primordial black holes produced in the very early universe) exist \cite{carrol},\cite{celotti}. For black holes produced in gravitational collapse, $T$ behaves as a physical time coordinate since, as proper ($\tau$) and coordinate ($t$) times, it also is future directed. Nevertheless, eternal black holes are solutions of Einstein equations, and for them, as shown in section 1, $T$ looses physical character in region $\bar{U}$.

\section*{Acknowledgment}

This work was partially supported by the project PAPIIT IN105413, DGAPA, UNAM.

%
%
%
%
%
%
%
%
%
%
%

\bibliographystyle{unsrt}
\bibliography{sample}

\vspace{\fill}
\begin{center}
\line(1,0){350}
\end{center}

e-mails: $^1$brauer@ciencias.unam.mx, $^2$hugo.camargo@correo.nucleares.unam.mx, 

$^3$socolovs@nucleares.unam.mx

\end{document}